\documentclass[aps, prd, reprint, nofootinbib, longbibliography, superscriptaddress, floatfix, preprintnumbers]{revtex4-2}

\usepackage{xurl}
\usepackage{hyperref}
\usepackage{amsmath}
\usepackage{amsfonts}
\usepackage{amssymb}
\usepackage{color}
\usepackage[utf8]{inputenc}
\usepackage{graphicx}
\usepackage{microtype}
\usepackage{siunitx}
\usepackage{soul}
\usepackage{longtable}
\usepackage{bm}
\usepackage{xspace,multirow }
\usepackage{natbib}
\usepackage[normalem]{ulem}
\usepackage[table]{xcolor}

\definecolor{tab-blue}{RGB}{0, 107, 164}
\hypersetup{colorlinks,linkcolor={blue},citecolor={teal},urlcolor={tab-blue}}

\newcommand{\gr}{$\gamma$-ray\xspace}
\newcommand{\grs}{$\gamma$\,rays\xspace}

\newcommand{\blj}{\texttt{BHJet}\xspace}

\newcommand{\mkn}{Markarian~421\xspace}
\newcommand{\mrk}{Markarian~421\xspace}
\newcommand{\txs}{TXS~0506+056\xspace}

\newcommand{\rg}{\,R$_g$\xspace}

\newcommand{\zdiss}{\,$z_{\rm diss}$\xspace}
\newcommand{\fsc}{\,$f_{\rm sc}$\xspace}
\newcommand{\sigmaDMe}{\,$\sigma_\text{DM-e}$\xspace}
\newcommand{\ljet}{\,$L_{\rm jet}$\xspace}

\newcommand{\INFN}{INFN - Sezione di Napoli, Complesso Universitario Monte Sant'Angelo, I-80126 Napoli, Italy}
\newcommand{\SSM}{Scuola Superiore Meridionale, Via Mezzocannone 4, I-80138 Napoli, Italy}

\begin{document}

\begin{flushleft}
LAPTH-010/25
\end{flushleft}

\title{Cosmic-ray cooling by dark matter in astrophysical jets}
\author{Dimitrios Kantzas}\email{dimitrios.kantzas@nyu.edu}
\affiliation{LAPTh, CNRS,  USMB, F-74940 Annecy, France}
\affiliation{Center for Astrophysics and Space Science (CASS), New York University Abu Dhabi, PO Box 129188, Abu Dhabi, UAE
}
\author{Francesca Calore}\email{calore@lapth.cnrs.fr}
\affiliation{LAPTh, CNRS,  USMB, F-74940 Annecy, France}

\author{Marco Chianese}
\email{m.chianese@ssmeridionale.it}
\affiliation{\SSM}
\affiliation{\INFN}

\begin{abstract}
Astrophysical jets from powerful active galactic nuclei (AGN) have recently been proposed as promising probes of dark matter (DM) in the sub-GeV mass range. AGN launch relativistic jets that accelerate cosmic rays (CRs) to very high energies, which can then interact with their surroundings and produce multiwavelength (MW) emission spanning from radio frequencies to TeV $\gamma$ rays. If DM consists of light particles, their interactions with CRs could lead to an additional cooling mechanism that modifies the expected MW emission.
In this work, we analyse the MW spectrum of Markarian 421, a well-studied AGN, using a multizone leptonic jet model that includes the interactions between CR electrons and DM particles. For the first time, we account for the uncertainties in the astrophysical jet dynamics, which have been previously neglected when constraining the CR-DM interactions.
By fitting simultaneously jet parameters and DM-electrons interactions, we use the MW data from \mkn to set constraints on the DM-induced CR cooling.
We obtain 5$\sigma$ upper limit $\sigma_\text{DM-e} \lesssim 1 \times 10^{-34}~\text{cm}^2$ for a DM mass of $1~{\rm MeV}$. We demonstrate that this is about a factor of 2--10 stronger than traditional approaches depending on DM mass. This improvement originates from having indeed considered the full multi-wavelength emission from the source, instead if a simplified approach. Properly accounting for degeneracies between jet dynamics and DM interactions 
is also key to deriving robust constraints on DM interactions. 
\end{abstract}

\maketitle

\section{Introduction}\label{sec:intro}

The existence of dark matter (DM) in the Universe is well established, but its nature remains elusive~\cite{Bertone:2004pz, Bertone:2016nfn, Kahlhoefer:2017dnp, Planck:2018vyg, Billard:2021uyg, AlvesBatista:2021eeu}. The sub-GeV mass regime is largely unexplored~\cite{Billard:2021uyg}, as current direct detection experiments are more sensitive to DM particles with masses above the GeV scale, owing to their limited sensitivity at low nuclear recoil energies. Various indirect detection strategies have been proposed to probe sub-GeV DM particles for masses still allowed by both Big Bang Nucleosynthesis (BBN) and direct-detection bounds~\cite{Knapen:2017xzo, CRESST:2015txj, Emken:2018run, CRESST:2017ues, Kouvaris:2016afs, Dolan:2017xbu}. One possibility involves the interaction of DM particles with cosmic rays (CRs), which can result in two key effects. First, collisions between DM particles and CRs accelerated in astrophysical environments can produce a population of boosted DM particles, which may be detectable at the Earth~\cite{Bringmann:2018cvk, Ema:2018bih, Giudice:2017zke, Bondarenko:2019vrb, Cappiello:2019qsw, Alvey:2019zaa, Dent:2019krz, Berger:2019ttc, Wang:2019jtk, Ema:2020ulo, Guo:2020drq, Jho:2020sku, Dent:2020syp, Wang:2021jic, Granelli:2022ysi, Bell:2021xff, Feng:2021hyz, Das:2021lcr, Xia:2021vbz, Xia:2022tid, Super-Kamiokande:2017dch, PROSPECT:2021awi, PandaX-II:2021kai, CDEX:2022fig, Maity:2022exk, DeRomeri:2023ytt, Xia:2024ryt, Cappiello:2024acu, Ghosh:2024dqw, Gustafson:2025dff, Li:2025zwg, Chen:2025twc, Jeesun:2025gzt}. Second, such interactions could act as a cooling mechanism for CRs, thereby perturbing their non-thermal emission~\cite{Cappiello:2018hsu, Herrera:2023nww, Gustafson:2024aom, Mishra:2025juk}. In the inelastic regime, they can also lead to the cascade production of secondary particles such as \grs and neutrinos~\cite{Gorchtein:2010xa, Ambrosone:2022mvk, Lu:2023aar, DeMarchi:2024riu, DeMarchi:2025xag}.
These secondary particles may compete against the products of the standard astrophysical particle-emission processes, making the identification of a DM signals challenging to disentangle from purely astrophysical signatures.

Active galactic nuclei (AGNs) 
are black holes (BHs) known to launch relativistic outflows that remain collimated over kpc distances. These outflows, known as \textit{jets}, carry enough power to accelerate CRs to high energies, possibly beyond 1~PeV. The precise composition of CRs is still unknown, but both leptonic and hadronic scenarios are possible. Leptonic CRs can, in fact, explain the majority of the multiwavelength (MW) spectrum emitted for most sources, from radio to TeV \grs~\cite{Ghisellini:2009fj, Ghisellini:2014pwa, Fermi-LAT:2021ykq}. In some cases, hadronic CRs are accelerated in jets and the interactions with the ambient medium lead to the formation of secondary particles; the stable ones are pairs of electrons and positrons, neutrinos and \grs from neutral pion decay~\cite{Petropoulou:2015upa, Cerruti:2014iwa}. Since the electromagnetic emission fails to distinguish the two scenarios, the neutrino emission represents the smoking gun for hadronic acceleration in AGN jets~\cite{IceCube:2018cha}.

DM is present in galaxies according to astrophysical observations and simulations~\cite{Iocco:2015xga, Werhahn:2021bal}. Interestingly, in the vicinity of massive BHs at the center of galaxies, the DM density is expected to be enhanced. The adiabatic growth of BHs leads to a growth of the DM density, commonly referred to as DM spike~\cite{Gondolo:1999ef}. The presence of the DM spike makes AGNs as ideal targets for the investigation of boosted DM signals and CR cooling due to CR-DM collisions. 
Previous works have analysed the expected detection of boosted DM signals considering \txs and BL~Lacerta~\cite{Wang:2021jic, Granelli:2022ysi}. Both these sources are well-studied AGNs, and the former in particular is the first blazar (an AGN launching a jet along the line of sight) correlated to an astrophysical neutrino. Refs.~\cite{Herrera:2023nww, Gustafson:2024aom, Mishra:2025juk} have examined the CR cooling from DM-CR collisions for \txs and NGC~1068, a Seyfert galaxy correlated to a strong neutrino emission, to set bounds on the CR-DM cross-section.
Such studies, however, lack a consistent investigation of the impact of the uncertainties of the dynamical properties of the astrophysical source on the constraints in the DM parameter space. On the other hand, Ref.~\cite{Ambrosone:2022mvk} has performed a more consistent statistical analysis of the \gr flux of M82, a well-studied star forming galaxy, to examine the effects of the CR-DM interactions on the emitted spectrum. This approach is extremely promising to reveal a spectral deviation from the standard one for the case of blazars, especially when accompanied by a proper jet modelling.

In this work, we examine in a statistically self-consistent framework the effect of the uncertainty of the kinematics of astrophysical blazar jets in constraining the nature of the DM particles. We use in particular a multi-zone jet model developed to reproduce and predict the electromagnetic flux covering the entire MW spectrum from radio to very high energy \grs~\cite{Lucchini:2021scp}. We further develop this jet model (henceforth, we refer to the purely astrophysical case as \blj) to include the CR electron cooling due to CR-DM collisions and examine the jet dynamical quantities that are most degenerate with a DM signal. We analyse in particular the emitted spectrum of the jets of \mkn, one of the brightest and nearby blazars. 
\mrk is located at 1.22~Mpc at redshift 0.0308 from the Earth~\cite{Sbarufatti:2005kk}. The mass of the BH is $M_{\rm BH} = (1.9 \pm 1.2) \times 10^8~M_\odot$~\cite{Barth:2002ac}, which corresponds to a Schwarzschild radius of ${R_g} = GM_{\rm BH}/c^2= 1.8\times10^{-8}~{\rm kpc}$.
Using the simultaneous and MW spectrum from radio to TeV \grs, as presented by Ref.~\cite{ARGO-YBJ:2015qiq}, we examine the parameter degeneracies and place constraints on the DM-electron cross section.  

The paper is organized as follows. In Sec.~\ref{Sec: Jet model}, we present the jet model, and in Sec.~\ref{Sec: DM section} we discuss the DM density profile and the CR cooling due to DM particles. In Sec.~\ref{Sec: method} we discuss the methodology we follow, and in Sec.~\ref{Sec: Results} we present the results of our analysis. Finally, we discuss our findings and conclude in Sec.~\ref{Sec: conclusions}.

\section{Cosmic-ray acceleration and cooling in jets}\label{Sec: Jet model}

To capture the CR acceleration along the jet, as well as the jet launching and evolution, we use \blj, a semi-analytical, multi-zone jet model~\cite{Markoff:2002xs, Markoff:2005ht, Lucchini:2021scp}.
We assume that the jets are launched close to the BH at distance $2\, R_g$, and they extend conically forming 100 jet segments along the jet axis $z$. Each jet segment has cross-sectional radius $R(z)$ and height 2$R(z)$, and it is described by its particle number density $n(z)$ and its magnetic field strength $B(z)$. 
The jet base has a cross-sectional radius $R_0$, and an equal number density of electrons and protons. The protons carry the bulk mass of the jet, whereas electrons form a thermal, relativistic, Maxwell-J\"uttner distribution that peaks at $511\,$keV. 
The total power of the jet base \ljet is distributed between the particles and the magnetic field.

At distance \zdiss 
from the jet base, magnetic energy is dissipated into non-thermal particle energy.
In particular, we assume that this jet segment has a magnetization given by $B^2/(4\pi n  m_p) = 0.02$, which represents the ratio between the magnetic field energy density and the rest mass energy density.

A fraction $f_e$ of the thermal electrons gets accelerated to non-thermal energies.
The total number density of the non-thermal electrons is $n_{\rm nth}(E_e,z) = f_e \, n_e(E_e,z)$, where $n_e(E_e,z)$ is the total electron number density at distance $z$ along the jet. These electrons are distributed in a non-thermal power law following the steady-state transport equation
\begin{equation}
\label{eq:steady-state}
    \dfrac{n_{\rm nth}(E_e,z)}{t_{\rm total}} \simeq  Q(E_e,z)\,.
\end{equation}
Here, $t_{\rm total}^{-1} = \sum t_i(E_e,z)^{-1}$ is the sum of the loss rate per physical process $i$, and 
$Q(E_e,z) = Q_0(z)\, q(E_e)$ is the injected rate of non-thermal electrons per energy $E_e$, where $Q_0(z)$ is the normalization and $q(E_e) \propto E_e^{-p}\,\exp({-E_e/E_{e, \rm max})}$ with $p=2$ and $E_{e, \rm max}$ is the maximum attainable energy for each jet segment beyond \zdiss. 
We assume $p=2$ for every jet segment \citep{Hovatta:2014koa, Kravchenko:2025abc}, and we calculate $E_{e, \rm max}$ by comparing the characteristic timescale of acceleration to the one for losses, including synchrotron, inverse Compton (IC) scattering, and escape. Namely,
\begin{eqnarray}
    \label{eq: Emax calculation from timescales}
    t_{\rm acc}^{-1}(E_{e,\rm max}) &=& t_{\rm syn}^{-1}(E_{e,\rm max}) + t_{\rm IC}^{-1}(E_{e,\rm max}) \nonumber \\ &&+  t_{\rm esc}^{-1}(E_{e,\rm max}) \,,
\end{eqnarray}
with
\begin{eqnarray}
    t_{\rm acc} (E_e) &=&  \dfrac{4E_e}{3 f_{\rm sc} B}\,,
    \\
    t_{\rm syn} (E_e) &=& \dfrac{6\pi {m_e^2 } }{\sigma_T B^2 E_e \beta_e^2}\,,\\
    t_{\rm IC} (E_e) &=& t_{\rm syn}(E_e) \dfrac{U_B}{U_{\rm rad}}\,, \\
    t_{\rm esc} (E_e) &=& \dfrac{R}{\beta_e}\,. 
\end{eqnarray}
In the above equations, $f_{\rm sc}$ is the acceleration efficiency of a diffusive shock-type acceleration~\cite{jokipii1987rate}, which we assume to be constant at \fsc$=10^{-6}$ based on previous applications of \blj \citep{Connors2016, Lucchini:2018rju, Lucchini:2021scp}
$\beta_e$ is the electron velocity, 
$U_B=B^2/8\pi$ is the magnetic-field energy density, 
$U_{\rm rad}$ is the photon-field energy density,
and $R$ is the cross-sectional radius of the jet at distance $z$.
The rest of the parameters are constants: $m_e$ is the electron rest mass and $\sigma_T$ is the Thomson cross section. 

Having calculated the non-thermal electron distribution, we derive the synchrotron radiation for each jet segment. Both expressions for the emissivity and the absorption are those of Ref.~\cite{Blumenthal:1970gc}, where we assume cylindrical geometry for the jet segments~\cite{Lucchini:2021scp}. For the IC scattering, we use the IC kernel given in Ref.~\cite{Blumenthal:1970gc}, where the target photons are those produced due to synchrotron emission. In the following, we only examine the case of \mkn where the \gr emission is dominated by leptonic processes. Hence, we neglect any hadronic radiation that could act as target for IC, as well as any external photon field. These two photon targets may be important in other source types. 

Before discussing the fit of \mkn MW spectrum, we first identify the key parameters that significantly affect it. Having explored different initial conditions and different jet dynamical quantities, we conclude that the most impactful parameters modifying the MW spectrum are the jet base injected power \ljet, the radius at the jet base $R_0$ and the location of the energy dissipation where particle acceleration kicks in \zdiss. The last two parameters dictate the shape of the jet and hence its emission. 
We note that more parameters can alter the emitted spectrum, for instance the power-law index of the non-thermal particles. We discuss later on the impact of this specific parameter on our final results.

\begin{figure}[t!]
    \centering
    \includegraphics[width=1.\linewidth]{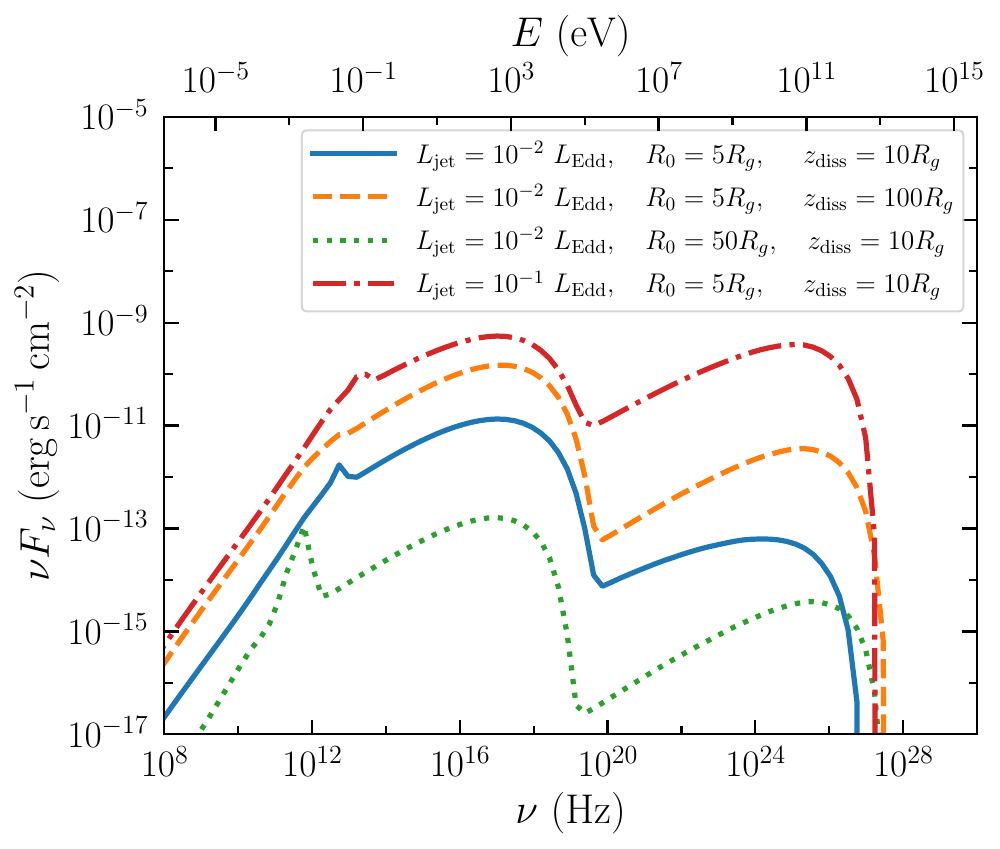}
    \caption{The MW emitted energy flux $\nu F_{\nu}$ of a \mkn-like blazar for different injected jet power \ljet, jet base radius $R_0$, and particle acceleration region \zdiss, as indicated in the legend. We assume a magnetization at \zdiss of 0.02 and a power-law index of the non-thermal electrons $p=2$.} 
    \label{fig:benchmark}   
\end{figure}

In Fig.~\ref{fig:benchmark}, we show the predicted energy flux $\nu F_{\nu}$ as a function  of the photon frequency $\nu$, in case of four difference choices for the three parameters.
The jets propagate with a bulk Lorentz factor of 25~\cite{ARGO-YBJ:2015qiq}.
When the dissipation region is at $z_{\rm diss} = 100\,{\rm R_g}$ (dashed orange line) instead of 10\rg (solid blue line), the magnetic energy density is increased for a constant magnetization, and hence the synchrotron luminosity increases.
When the jet base is larger, the size of the dissipation region increases as well. As a result, for a given magnetization, the magnetic energy density decreases, leading to lower synchrotron emission (dotted green line compared to solid blue line). 
Finally, when the injected power at the jet base \ljet is ten times larger, more energy is distributed in both the particles and the magnetic fields eventually leading to stronger total non-thermal emission (dotted-dashed red line compared to solid blue line). The spectral feature at $\sim 10^{12}~{\rm Hz}$ is the synchrotron emission of the thermal electrons at the jet base.

\section{Dark matter and cosmic rays}\label{Sec: DM section}

\subsection{DM profile and spike}

The radial DM density profile, $\rho_{\rm DM}(r)$, is given by the expression (see e.g., Ref.~\cite{Lacroix:2016qpq}):
\begin{equation}\label{equation: dark matter profile}
    \rho_{\rm DM}(r) = 
    \begin{cases}
        0 & r< 4R_g \\
        \dfrac{\rho_{\rm sp}(r)\rho_{\rm core}}{\rho_{\rm sp}(r)+\rho_{\rm core}} & 4R_g\leq r \leq R_{\rm sp} \\
        \rho_s\left( \dfrac{r}{r_s} \right)^{-\gamma} \left( 1+\dfrac{r}{r_s} \right)^{\gamma-3}  & R_{\rm sp}< r \leq r_{200}
    \end{cases}
\end{equation}
In general, towards the BH center, the density profile converges to the constant value, $\rho_{\rm core} = m_{\rm DM} /( \left< \sigma v \right> t_{\rm BH} )$, which depends on the BH growth timescale $t_{\rm BH}$, the DM mass $m_{\rm DM}$, and the velocity-averaged DM self-annihilation cross section $\left< \sigma v \right>$. Conversely, it is generally assumed that no DM is present within a radius of $4R_g$ from the BH. The spike density takes the expression~\cite{Gondolo:1999ef}:
\begin{equation}
    \rho_\text{sp} (r) = \rho_s \left( 1-\frac{4R_\text{g}}{r} \right)^3 \left(\frac{R_\text{sp}}{r_s} \right ) ^{-\gamma}  \left(\frac{ R_\text{sp}}{r} \right)^{\gamma_\text{sp}},
\end{equation}
with slope $\gamma_\text{sp} = (9-2\gamma)/(4-\gamma)$ and extension
\begin{equation}\label{eq:Rspike}
R_{\rm sp} = \alpha_\gamma r_s \left( \frac{M_{\rm BH}}{\rho_s r_s^3} \right) ^{1/(3-\gamma)}\,,
\end{equation}
where $\alpha_\gamma = 0.122$ for $\gamma=1$. Assuming $\gamma=1$, the DM profile beyond the spike radius evolves radially as a Navarro-Frenk-White (NFW) profile~\cite{Navarro:1995iw, Navarro:1996gj}. Therefore, in addition to the quantity $\rho_{\rm core}$, the total DM profile is fully determined by the BH mass $M_{\rm BH}$ (which is often measured), the scale radius $r_s$ and the normalization $\rho_s$.

We compute the quantities $r_s$ and $\rho_s$ as follows. We first determine the total DM halo mass $M_{\rm halo}$\footnote{This is also defined here as $M_{200}$, and matches the definition of Ref.~\cite{Marasco:2021pkl}.} 
from the $M_{\rm BH}-M_{\rm halo}$ relation obtained in Ref.~\cite{Marasco:2021pkl} through the observation of 55 nearby galaxies. 
Given the total DM halo mass, we compute the value of the radius $r_{200}$ -- defined as the radius within which the average DM density is 200 times the critical value $\rho_{\rm cr} = 8.5\times 10^{-27}\,\rm kg\,m^{-3}$ -- from the relation~\cite{Boudaud:2021irr}:
\begin{equation}
    M_{\rm halo} = \frac{4\pi}{3} r_{200}^3 \, 200\rho_{\rm cr}\,.
\end{equation}
Then, we obtain the radius $r_s$ and its uncertainty using the results for the concentration parameter $c_{200} (M_{\rm halo}) = r_{200}/r_s$ obtained in Ref.~\cite{Asgari:2023mej} according to the simulations at redshift $z=0$ discussed in Ref.~\cite{Ishiyama:2020vao}.
Finally, the normalization of the profile $\rho_s$ can be determined by solving the equation:
\begin{equation}\label{eq: DM halo mass}
    M_{\rm halo} = 4\pi \int_{4R_g} ^{r_{200}} r'^2 \rho_{\rm DM}(r') \,{\rm d}r' \,.
\end{equation}
We stress that this procedure differs from what was done in the previous literature (e.g. in Refs.~\cite{Lacroix:2015lxa, Lacroix:2016qpq, Wang:2021jic, Granelli:2022ysi, Herrera:2023nww}), but it is, in our opinion, a more self-consistent treatment to derive the DM halo parameters. Further details on the computation of the density profile, as well as a comparison with previous derivations, are provided in Appendix~\ref{app: DM profile calculations}. For the specific case of \mkn, we find the DM halo parameters reported in  Tab.~\ref{tab:DMparameters}. For the entire analysis we assume non-annihilating DM particles $(\left<\sigma v\right>/ {\rm (cm^3\, s^{-1})}  = 0 )$, which leads to stronger bounds on the CR-DM interaction. The presence of self-annihilating DM particles would result in a reduced DM density of the spike in the regions close to the jet launching and particle acceleration, hence diminishing the target density of CR-DM collisions and weakening the final bounds. 
We will discuss the case of non-zero annihilation cross section in Sec.~\ref{Sec: Results}.
\begin{table}[t!]
    \centering
    \begin{tabular}{lc}
        Parameter & Value \\
        \hline 
        $M_{\rm halo}/M_\odot$ & $5.7 \pm 3.3\times 10^{12}$ \\
         $r_{200}/$kpc & $377\pm 74$ \\
         $c_{200}$ & $6.21\pm 0.39$\\
         $r_s/$kpc &$ 60.81 \pm 12.58$ \\
         $\rho_s/(M_{\odot}\, \rm kpc^{-3})$ & $1.8_{-1.3}^{+3.1} \times 10^6$
    \end{tabular}
    \caption{DM halo parameters and their 1$\sigma$ uncertainty that define the radial DM profile described in Eq.~\eqref{equation: dark matter profile} for \mkn. We have considered $\left< \sigma v \right> / {\rm (cm^3\, s^{-1})} = 0$ in the computation of $r_s$ and $\rho_s$.}
    \label{tab:DMparameters}
\end{table}

\subsection{CR cooling due to DM}

The interaction between the accelerated CR electrons and DM particles leads to an additional energy-loss process which modifies the transport of CR electrons along the jets. The timescale in which CR-DM collisions occur is given by \cite{Ambrosone:2022mvk}
\begin{equation} \label{eq:tDM-e}
    t_\text{DM-e} (E_e) = \left[ - \dfrac{1}{E_e} \left(\dfrac{{\rm d} E_e}{{\rm d}t}\right)_\text{DM-e} \right]^{-1}\,, 
\end{equation}
with the energy loss of CR electrons defined as
\begin{equation}
    \left(\dfrac{{\rm d} E_e}{{\rm d}t}\right)_\text{DM-e} = -\frac{\rho_{\rm DM}}{m_{\rm DM}} \int _0 ^{T_{\rm DM}^{\rm max}} T_{\rm DM} \frac{{\rm d}\sigma_{\rm el}}{{\rm d}T_{\rm DM}}\,{\rm d}T_{\rm DM}\,.
\end{equation}
The quantity $T^{\rm max}_{\rm DM}$ represents the maximal allowed value for the kinetic energy of DM particles $T_{\rm DM}$ in a collision with an electron with kinetic energy $T_e = E_e -  m_e$, namely \cite{Bringmann:2018cvk,Emken:2018run}
\begin{equation}
    T^{\rm max}_{\rm DM} = \frac{ 2T_e^2 + 4{m_e}T_e }{m_{\rm DM}} \left[ \left(1 + \frac{{ m_e}}{m_{\rm DM}} \right)^2 + \frac{2T_e}{m_{\rm DM}} \right]^{-1}\,.
\end{equation}
The differential elastic DM-electron cross section takes the expression \cite{Bondarenko:2019vrb,Ema:2020ulo}
\begin{equation}\label{Eq: DM electron cross section}
    \frac{{\rm d}\sigma_{\rm el}}{{\rm d}T_{\rm DM}} = \frac{\sigma_\text{DM-e}}{T^{\rm max}_{\rm DM}}\frac{1}{16 \mu^2_\text{DM-e}s} (q^2 + 4{m_e^2}) ( q^2 + 4m^2_{\rm DM}) \,,
\end{equation}
where $\sigma_\text{DM-e}$ is the DM-electron cross section at zero center-of-mass momentum, $\mu_\text{DM-e}$ is the reduced mass of DM and electron, $s = m_{\rm DM}^2 + { m^2_e} + 2E_em_{\rm DM}$ is the center-of-mass energy, and $q^2 = 2m_{\rm DM}T_{\rm DM}$ is the momentum transfer. Eq.~\eqref{Eq: DM electron cross section} holds for heavy mediators, namely for $q^2 \ll m_s^2$ with $m_s$ being the (heavy) mediator rest mass \cite{Bondarenko:2019vrb}. Since the maximum kinetic energy of the DM particle is constrained by the maximum energy of the CR electrons, the above inequality holds for our study as long as $m_s \gg 10~{\rm GeV}$.

To account for the CR cooling due to CR-DM collisions, we include the CR-DM timescale in the calculation of the non-thermal electron distribution in Eq.~\eqref{eq:steady-state} as well as in the one of the $E_{e,\rm max}$ in Eq.~\eqref{eq: Emax calculation from timescales}. The CR-DM cooling, indeed, affects not only the maximum energy of the accelerated CRs, but also the normalization of the CR power-law distribution. The emitted radiation can consequently differ significantly from the standard one, as we discuss below.

\begin{figure}[t!]
    \includegraphics[width=1.\columnwidth]{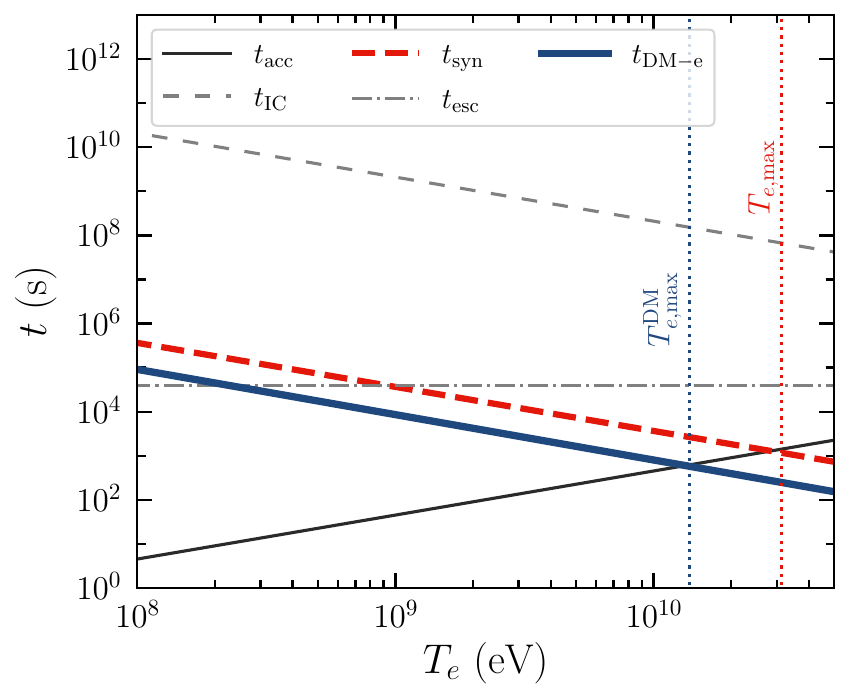}
    \caption{\label{fig:timescales} The characteristic timescales of electron acceleration and losses as a function of the electron kinetic energy, as they occur in the jet segment at \zdiss. The solid thin black line represents the acceleration timescale, while standard astrophysical losses are given by the dot-dashed grey line (escape), the dashed grey line (IC), and the dashed red line (synchrotron). The thick solid blue line corresponds to the timescale of CR-DM collisions (see Eq.~\eqref{eq:tDM-e}) assuming $m_{\rm DM} = 1~{\rm MeV}$ and $\sigma_\text{DM-e}=10^{-34}~{\rm cm^2}$. The intersection between the acceleration and the fastest cooling process (lowest cooling timescale) gives the maximum attainable kinetic energy of the electrons $T_{e,{\rm max}}$, which is indicated by the vertical lines for the case with and without DM cooling.}
\end{figure}

The CR-DM collisions lead to less energetic CR electrons, whose maximum energy is reduced compared to the scenario without DM.
Fig.~\ref{fig:timescales} shows the cooling timescales and the particle acceleration timescale for the case with and without DM cooling. More precisely, 
we plot the synchrotron emission (dashed red line), the IC scattering (dashed grey line) and the escape timescale (dot-dashed grey line) for the jet segment at \zdiss. In the case of DM cooling, we assume $\sigma_\text{DM-e}=10^{-34}\rm\,cm^2$ and $m_{\rm DM} = 1~{\rm MeV}$ (solid blue line). In the case that there is DM cooling, the maximum electron energy drops slightly above 10\,GeV, as shown by the leftmost vertical line. 

\section{Fitting method and dark matter upper limits}\label{Sec: method}

We explore the 3D parameter space defined by the astrophysical parameters $\theta_{\rm astro} = \{L_{\rm jet}, R_0,\,z_{\rm diss}\}$ to search for the best fit of the \blj calculated emission to the steady state MW observations of \mkn~\cite{ARGO-YBJ:2015qiq}. 
We choose to constrain the jet base power between $10^{-2}~L_{\rm Edd}$ and $1.25~L_{\rm Edd}$, where $L_{\rm Edd} = 1.26\times 10^{38}\,(M_{\rm BH}/M_{\odot})\,\rm erg\,s^{-1}$, the jet base radius between $2R_g$ and $30R_g$, 
and the dissipation region to extend up to $55R_g$.
We follow a standard $\chi^2$ procedure, namely we minimize the function
\begin{equation}
    \chi ^2(\theta_{\rm astro}) = \sum _i \left( \frac{ \mathcal{O}_i  - \mathcal{E}_i(\theta_{\rm astro}) }{\sigma_i} \right)^2\,,
\end{equation}
where $\mathcal{O}$ is the observed flux at energy bin $i$, $\mathcal{E}$ is the expected flux, and $\sigma$ the uncertainty of the observation.
We focus on the data as collected during the steady state between 2008 August and 2009 June~\cite{ARGO-YBJ:2015qiq}. 
The dataset consists of simultaneous observations in the radio, infrared, X-ray and \gr bands~\cite{ARGO-YBJ:2015qiq}. The jet emission dominates all these wavelengths, with synchrotron emission peaking in the X-ray band and IC scattering in the \gr one. The X-ray emission, specifically, shows features of neutral photoelectric absorption in the interstellar emission and reflection effects, but their detailed study remains outside the scope of our analysis. To incorporate these phenomena in the uncertainty of our predicted flux, we force an increased systematic uncertainty of 20\% 
of the observed flux in the X-ray band. This assumption also facilitates the numerical estimation of the $\chi^2$ minimum, which is otherwise hindered by the large number of data points.

Then, we set the upper limits of \sigmaDMe by jointly fitting the astrophysical parameters and the CR-DM scattering cross section to the MW data of \mkn, following the same procedure as above. More specifically, we perform individual scans of the 4D phase space for different values of the mass $m_{\rm DM}$ of the DM particles. To gauge the difference from the null hypothesis, namely the purely astrophysical scenario, we calculate the quantity $\Delta \chi^2 = \chi^2 (\theta_{\rm astro},\,\sigma_\text{DM-e}; \,m_{\rm DM}) - \chi^2 (\theta_{\rm astro},\,0; \,m_{\rm DM})$, where the parameters $\theta_{\rm astro}$ are treated as nuisance parameters, and \sigmaDMe=0 means there is no DM effect regardless of the value of $m_\text{DM}$. We set bounds at $5\sigma$ confidence level requiring $\Delta \chi ^2 <23.6$ according to the Wilks theorem~\cite{Wilks:1938dza}. 

To make a direct comparison with the methodology adopted in previous analyses~\cite{Herrera:2023nww, Gustafson:2024aom, Mishra:2025juk}, albeit for other sources, we also compute the upper limits on \sigmaDMe by requiring that the CR-DM cooling timescale is 10 times faster than the total losses $t_{\rm total}$, namely, 
\begin{equation}\label{eq: timescale constraints}
    t_\text{DM-e} \lesssim 0.1\,t_{\rm total}\,.
\end{equation}
In the case of \mkn, the total timescales of losses is given by $t_{\rm total} \simeq  (t_{\rm syn}^{-1} + t_{\rm esc}^{-1}) ^{-1}$, as the IC cooling is negligible along all jet segments in our case (see Fig.~\ref{fig:timescales}). We emphasize once more that we calculate the cooling timescales and the radiation for every jet segment beyond \zdiss but the dominant radiative region is at \zdiss. In outer segments, where the radius is increased and the magnetic field is decreased, the maximal attainable energy is decreasing. 
Following Refs.~\cite{Herrera:2023nww, Gustafson:2024aom, Mishra:2025juk}, for this comparative calculation, we fix the jet parameters to the values obtained from the best fit without CR-DM interactions.

\section{Results}\label{Sec: Results}

\begin{figure}[t!]
    \includegraphics[width=1.\linewidth]{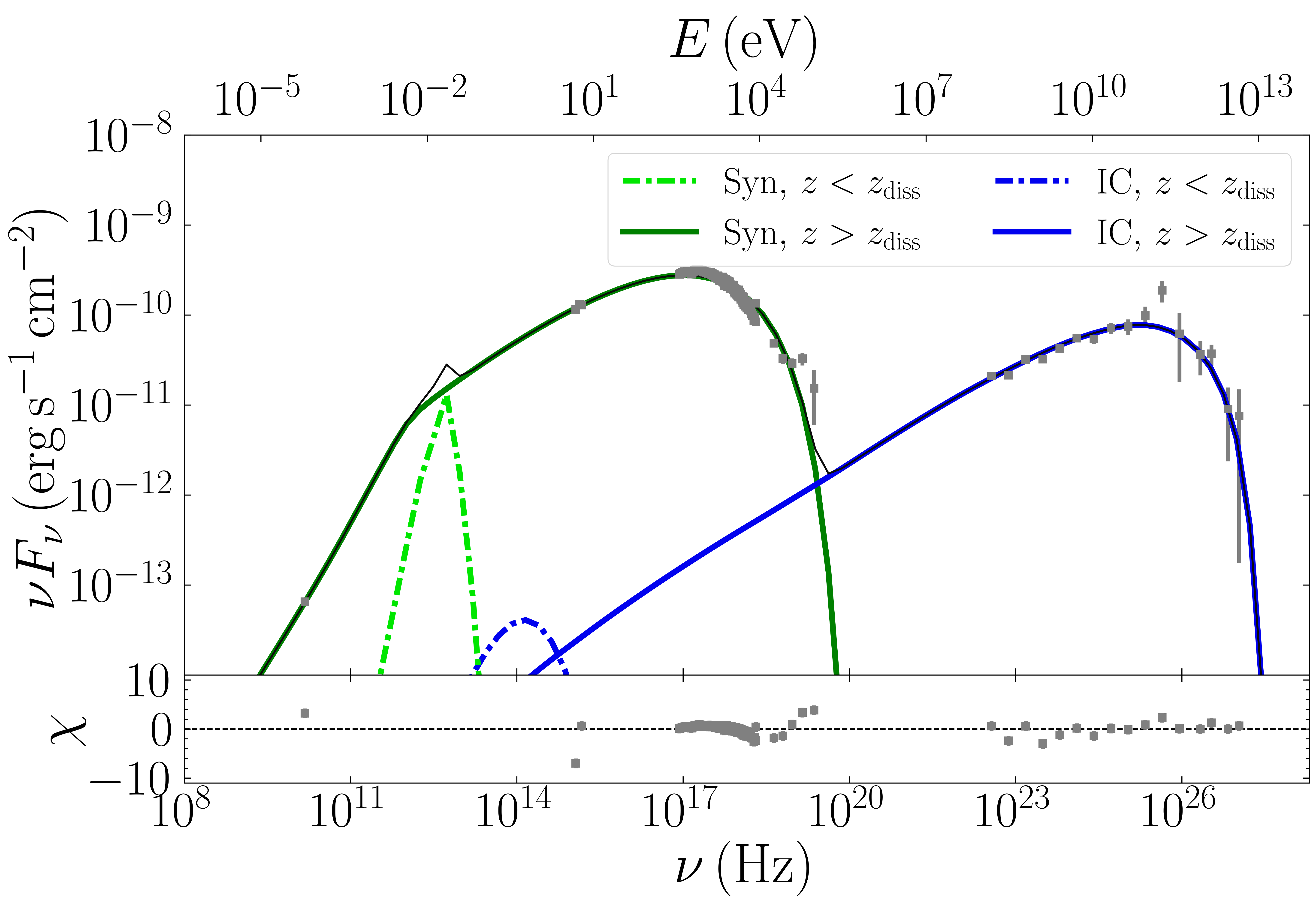} 
    \caption{\label{fig:astroBF} The spectral energy distribution of \mkn with the best fit of the total predicted emission to the MW dataset \cite{ARGO-YBJ:2015qiq}. The contribution of different processes and different jet regions are indicated in the legend, with the synchrotron emission (Syn) dominating the radio-to-X-ray spectrum, and the Inverse-Compton (IC) scattering explaining the \gr one. The bottom panel shows the residuals of the fit.}
\end{figure}

Fig.~\ref{fig:astroBF} shows the best fit of the calculated jet emission to the MW observations of \mkn.
We find the following best-fit parameters:
\begin{eqnarray}
    L_{\rm jet} &=& (99\pm1)\times 10^{-3}\,L_{\rm Edd}\,,\nonumber \\
    R_0 &=& (20.1 \pm 0.1) \,R_g \,,\\
    z_{\rm diss} &=& (37\pm 1)\,R_g \,.\nonumber
\end{eqnarray}
Similarly to previous works, we confirm that the IC scattering explains the GeV-to-TeV observations~\cite{1996A&AS..120C.503G, Mastichiadis:1996yi, Tavecchio:1998xw, 2001ApJ...554..725T, Konopelko:2003zr, Finke:2008pe, ARGO-YBJ:2011yuo,  Takahashi:2013lba, Petropoulou:2013lwa, Acciari:2013lwa, Abdo:2014vpa, Nilsson:2018tet}. Following the procedure described in Sec.~\ref{Sec: Jet model}, we find that the CR electrons are accelerated up to energies of 30\,GeV due to energy losses (see Fig.~\ref{fig:timescales}). In Fig.~\ref{fig:astroBF}, we also show the contribution of the thermal electrons at the jet base that peaks at $\sim 10^{12}\,$Hz.

\begin{figure}[t!]
    \includegraphics[width=1.05\columnwidth]{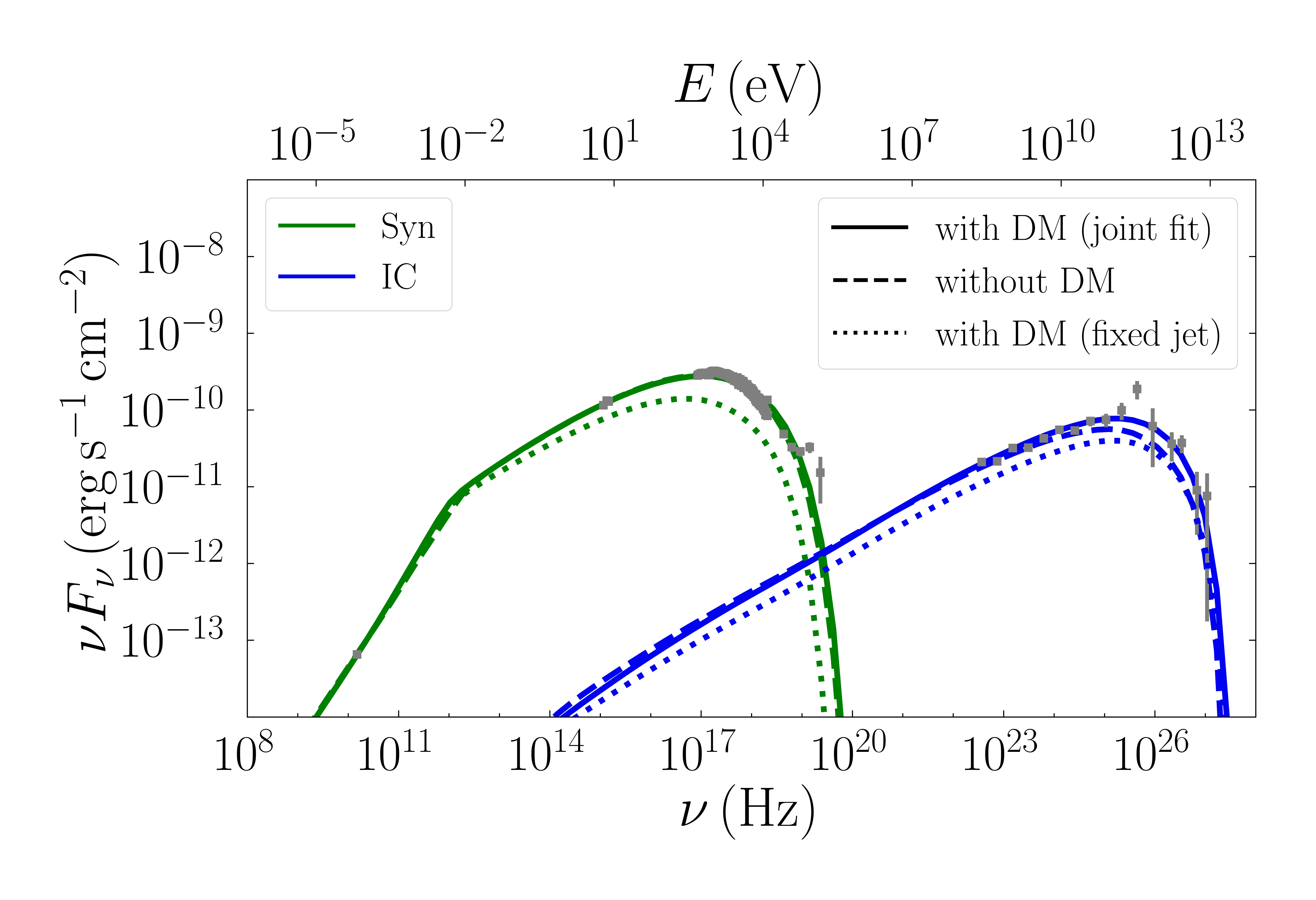}
    \caption{\label{fig: SED with DM}  Similar to Fig.~\ref{fig:astroBF}, but for the case where the emitted spectrum is modified due to CR cooling by elastic scattering on the DM particles, and jet modelling uncertainties are accounted for in the joint fit (solid line). The best astrophysical model (dashed line, labelled as \textit{without DM}) is shown for comparison. We also plot the case where the jet parameters remain the same as the astrophysical case and the DM cooling occurs with $m_{\rm DM} = $ 1 MeV and \sigmaDMe$=10^{-34}\, \rm cm^2$ (dotted line, labelled as \textit{fixed jet}). The green lines show the synchrotron component and the blue lines show the IC component.
}
\end{figure}

Fig.~\ref{fig: SED with DM} shows instead the emitted spectrum of \mkn considering the jets embedded in the DM spike with $m_{\rm DM} = 1\,\rm MeV$ and $\sigma_\text{DM-e}=10^{-34}\, \rm cm^{-2}$. 
Through the joint DM-jet fit, we can still obtain a MW spectrum compatible with the observed MW data (solid line). In this case, CR electrons are accelerated up to energies of 10~GeV (see Fig.~\ref{fig:timescales}), while the new best-fit jet parameters are
\ljet$=0.035~L_{\rm Edd}$, $R_0=4~R_g$, and $z_{\rm diss}=23~R_g$.

Compared with the best-fit without DM
(dashed line), both predicted spectra fit well the observational data with the main difference being in the cutoff of the emitted radiation, which has nonetheless larger errors compared to the rest of the spectral energy distribution data points. This difference in the emitted spectrum is due to the different maximum energy the CR electrons attain and the normalization of the non-thermal electrons that depends on the DM cooling. 

For illustrative purposes, we also overlay
the MW emission obtained when
fixing the jet parameters to the best-fit values for the astrophysical-only model (same as in Fig.~\ref{fig:astroBF}) and including the DM-CR interactions without re-fitting (dotted line). In this case, the effect of DM cooling is overestimated since there is no freedom of the astrophysical model, and degeneracies are not taken into account. 

From this plot, we can therefore conclude that the presence of DM particles can significantly affect particle acceleration and subsequent non-thermal MW emission through elastic cooling, and that accounting for astrophysical uncertainties alters the overall effect of DM cooling to the spectral data. 

\begin{figure}[!h]
    \includegraphics[width=1\columnwidth]{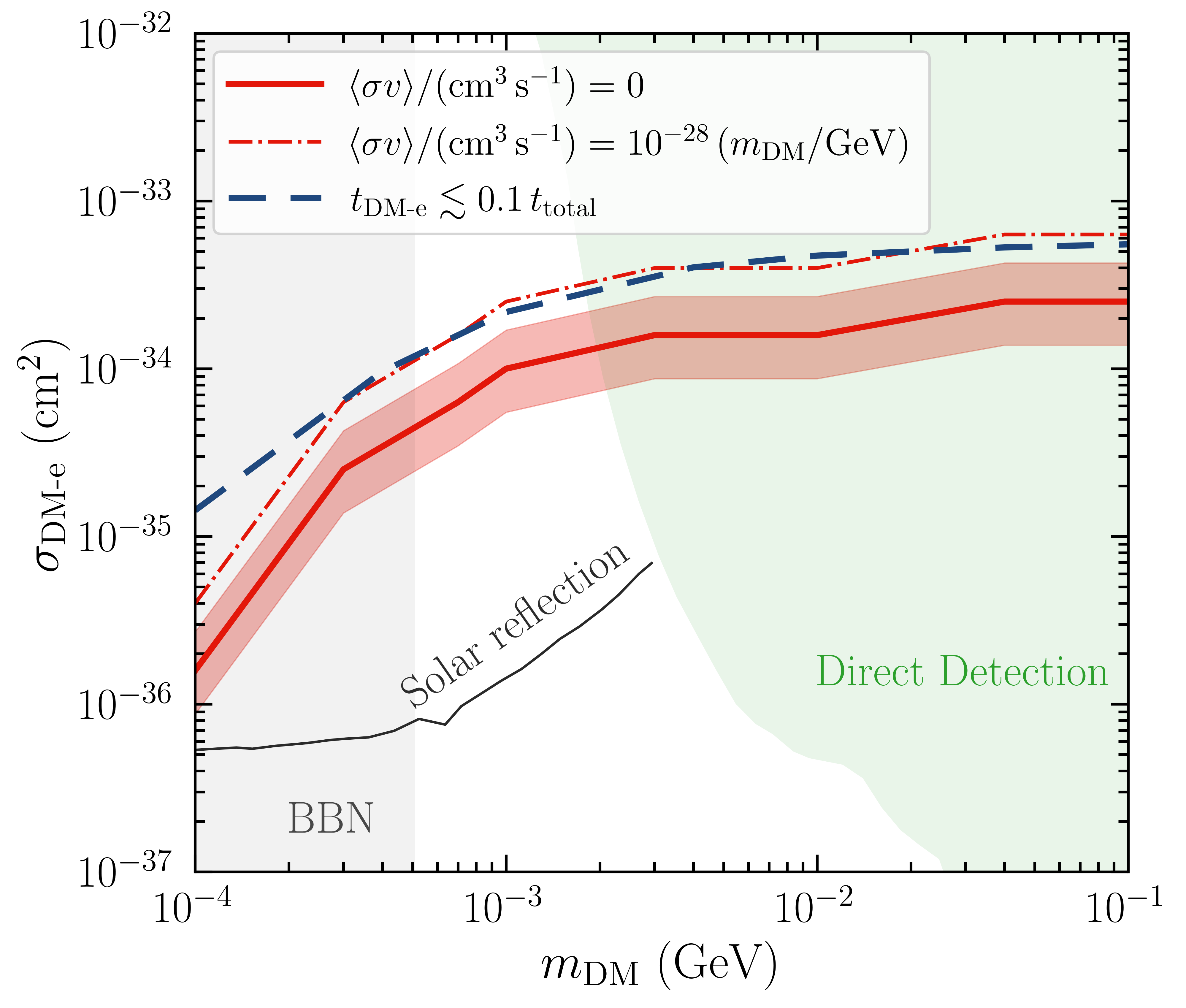} 
    \caption{\label{fig: DM phase space unconstrained} The upper limits of the cross section of DM-electrons interactions for different values of the mass of the light DM candidate. The light shaded areas are the constrained regions by other experiments and methods. In particular, the BBN \cite{Knapen:2017xzo}, and direct detection \cite{CRESST:2015txj, Emken:2018run, CRESST:2017ues, Kouvaris:2016afs, Dolan:2017xbu}. We also include upper bounds set by the solar reflection with thin solid black line \cite{An:2017ojc}. 
    The red lines represent the  5$\sigma$ upper limits from the joint fit to the MW emission of \mkn (this work). They show different
    DM self-annihilation scenarios as indicated in the legend. The red shaded band quantifies the uncertainty related to the DM distribution in \mkn.
    For comparison, we show the optimistic bounds (blue dashed line) set by following a very simplistic approach, namely by constraining the timescales as explained by Eq.~\eqref{eq: timescale constraints}.}
\end{figure}

In Fig.~\ref{fig: DM phase space unconstrained}, we present the upper limits on the DM–electron cross section for various DM particle masses. 
The bounds derived in this analysis are shown in red and blue colours, while previous constraints are indicated by the black line (solar reflection; \cite{An:2017ojc}), and shaded grey (BBN; \cite{Knapen:2017xzo}) and green (direct detection; \cite{CRESST:2015txj, Emken:2018run, CRESST:2017ues, Kouvaris:2016afs, Dolan:2017xbu}) regions.

The red limits corresponds to the main results of this work, derived through a joint fit of the jet dynamics and the DM-CR interactions. The solid line represents the case of non-self-annihilating DM ($\langle \sigma v \rangle / {\rm (cm^3\, s^{-1})}  = 0$), and the shaded region shows the upper bounds of \sigmaDMe accounting for the propagated uncertainties of $r_s$ and $\rho_s$ (see Table~\ref{tab:DMparameters}). Uncertainties from the DM distribution are no more than a factor of 1.7 for all masses.

The dot-dashed line corresponds to the more conservative upper limits in the case of self-annihilating DM with $\langle \sigma v \rangle / ({\rm cm^3\, s^{-1})}  = 10^{-28}\,(m_{\rm DM}/ {\rm GeV})$, a value that is in good agreement with recent constraints from X-ray observations with eROSITA~\cite{Balaji:2025afr}, and also similar to previous works~\cite{Granelli:2022ysi, Wang:2019jtk, Herrera:2023nww}. In this case, we obtain an upper bound of \sigmaDMe$=1.58\times10^{-34}\,\rm cm^2$, i.e.~$\sim2$ times weaker limits compared to the non-self-annihilating scenario.

We note that, for values of DM mass $m_{\rm DM} \gtrsim m_e$, the limits do not change significantly, and they almost stay constant. Integrating analytically Eq.~\eqref{Eq: DM electron cross section}, we indeed find that the CR-DM timescale scales as $t_\text{DM-e}\propto m_e^2/(\sigma_\text{DM-e}\,T_e)$, namely it does not depend on the mass of the DM particles. On the other hand, for $m_{\rm DM} \lesssim m_e$, from Eq.~\eqref{eq:tDM-e} we have that $t_{\rm DM-e} \propto m_{\rm DM}^2 / (\sigma_\text{DM-e} \, T_e)$, which explains the dependence of the constraints on the DM mass.

In Fig.~\ref{fig: DM phase space unconstrained}, we also include the upper limits derived by following the ``timescale'' approach as described in Sec.~\ref{Sec: method} using Eq.~\eqref{eq: timescale constraints}. 
The limit is \sigmaDMe = $2.4\times 10^{-34}\,\rm cm^{2}$ at 1\,MeV as shown by the dashed blue line. We see that this approach leads to weaker results despite neglecting the uncertainties of the jet parameters.
In fact, the number density of the non-thermal electrons, and consequently the synchrotron emission, scales as $n_{\rm nth} \propto t_\text{DM-e}$ (see Eq.~\ref{eq:steady-state}). In contrast, the inverse-Compton emission depends on $n_{\rm nth}^2$, which strengthens the resulting constraints in the mass regime $m_{\rm DM} \lesssim m_e$, making them more stringent than those obtained from the ``timescale'' approach.

\section{Discussion and Conclusions}\label{Sec: conclusions}

In this work, for the first time, we examine the effect of sub-GeV DM on CR acceleration in jets assessing the uncertainties of the jet modelling. 
We focused on the MW emission of Markarian 421. By capturing jet physics and the CR acceleration with a simple analytical framework, we find the contribution of the jet emission of \mkn to the MW spectrum for a purely astrophysical scenario. In this  scenario, a power law of non-thermal CR electrons is formed in the dissipation region \zdiss of the jets, where energy is dissipated to accelerate non-thermal particles \cite{Lucchini:2021scp}. 
The geometry and the jet power are the quantities that more strongly affect the emitted spectrum, hence, we performed a statistical analysis for three parameters, \ljet, $R_0$, and \zdiss. We find that a Poynting flux dominated jet that converts its magnetic energy to kinetic and non-thermal particles, can explain well the MW data of \mkn. We find that a non-thermal population of CR electrons with power-law index of 2 can explain well the radio observations. The \gr spectrum originates by IC scattering of synchrotron-produce electrons from the dissipation region of radius $20\,R_g$ at distance $z_{\rm diss} = 37\,R_g$ from the jet base, in agreement with previous works~\cite{1996A&AS..120C.503G,Mastichiadis:1996yi,Tavecchio:1998xw, 2001ApJ...554..725T, Konopelko:2003zr, Finke:2008pe, ARGO-YBJ:2011yuo,  Takahashi:2013lba, Petropoulou:2013lwa, Acciari:2013lwa, Abdo:2014vpa, Nilsson:2018tet}. That proves once more that the non-thermal emission of \mkn originates purely in leptonic processes and hadronic acceleration, which is more energy demanding, is not required. Moreover, the lack of  evidence of neutrino emission from the region of \mkn favors the leptonic scenario.  

We then examined the scenario where DM particles constrain the CR acceleration via elastic collisions. 
First, we revisit the DM profile around the galactic center of \mkn normalizing it by accounting for the entire DM halo (see Section~\ref{Sec: DM section}). We stress that we adopt a method in the normalization of the DM profile which is different from previous works, and yet, we believe, more consistent and precise. The difference between the peak of the calculated DM profile following our approach and the traditional ones is of the order of $\sim$ 1/50. Since the number density of DM particles scales linearly with the estimated CR cooling, the resulting upper limits in the CR-DM cross section may be overestimated by such a factor, cf.~Appendix~\ref{app: DM profile calculations}.  

We then perform a $\Delta \chi^2$ analysis where we fit for both the jet parameters \ljet, $R_0$, and \zdiss (similarly to the purely astrophysical scenario) and the CR-DM cross section \sigmaDMe for every different DM mass $m_{\rm DM}$ in the sub-GeV regime, as described in Sec.~\ref{Sec: method}.
Both scenarios, with and without CR cooling due to DM collisions, fit well the MW data with a reduced $\chi^2$ value close to 1. 

We notice that our astrophysical model cannot perfectly match the X-ray data that show  
features of neutral photoelectric absorption in the interstellar emission and reflection effects. 
We have checked that the significance of these best fits significantly reduces when we include a 20\% systematic uncertainty on X-ray data accounting for the astrophysical mis-modelling due to additional components. On the other hand, the upper limits we place on \sigmaDMe remain unaffected. Lower values of the uncertainty worsen the values of $\chi^2$ without altering the constraints of \sigmaDMe, hence no further modelling of the X-ray emission is required in our current setup.  

To fully assess the impact of our new methodology, which combines in a joint fit 
the jet dynamics and the DM interactions, we have compared our limits with the ones derived through the ``timescale approach'', as proposed in~\cite{Herrera:2023nww} albeit for another target. 
By adapting this approach to the case of \mkn, we demonstrate that fixing the CR-DM timescale to 10 times faster than the rest of the losses (see Eq.~\eqref{eq: timescale constraints}) leads to weaker constraints by a factor of 2-10 depending on the DM mass, and also fails to properly capture degeneracies with jet physics. We stress that the strength of our bounds comes from having considered the whole SED information and modeling components.

The constraints of \sigmaDMe are weaker compared to, for instance, searchers from Solar reflection~\cite{An:2017ojc}, and yet complementary to them.  
A step forward in our work will be the analysis of other jetted sources which are neutrino candidates, so to include hadronic acceleration \cite{Kantzas:2020mld, Kantzas:2023tzw, Dekker:2025vcg} and explore the effects of inelastic scattering between jet-accelerated CR and DM. 

Our proposed method represents a viable and promising way to constrain light DM in a window where the reach of direct 
detection experiments is still limited.
We have demonstrated that the inclusion of more accurate astrophysical models in the indirect DM searches is key to achieve robust results. Physical jet modelling in combination with a more extended coverage of the jet phase space is more crucial than ever:  with the up-coming high-resolution facilities, such as the Cherenkov Telescope Array Observatory (CTAO), very precise spectral determination for AGNs beyond 30\,GeV are expected, therefore enhancing the sensitivity to DM signatures in jets. 


\begin{acknowledgments}
We are grateful for the very fruitful comments of the
anonymous reviewer. We would like to thank Yoann Genolini for fruitful discussions on handling the data uncertainties, and Pierre Salati for discussions on the DM profiles. We would like to thank Pierre Aubert for his assistance on software acceleration, and the facilities offered by the Univ.
Savoie Mont Blanc - CNRS/IN2P3 MUST computing center.
DK and FC acknowledge funding from the French Programme d’investissements d’avenir through the Enigmass Labex, and from the ``Agence Nationale de la Recherche'', grant number ANR-19-CE310005-01 (PI: F. Calore). 
DK acknowledges support from Tamkeen under the NYU Abu Dhabi Research Institute grant CASS.
MC acknowledges support from the research project TAsP (Theoretical Astroparticle Physics) funded by the Istituto Nazionale di Fisica Nucleare (INFN).

\end{acknowledgments}

\section*{Data Availability}

The data are not publicly available. The data are available from the authors upon reasonable request.

\bibliography{bibliography_DMjets}

\appendix

\section{DM profile}\label{app: DM profile calculations}

\begin{figure}[t!]
    \includegraphics[width=1\columnwidth]{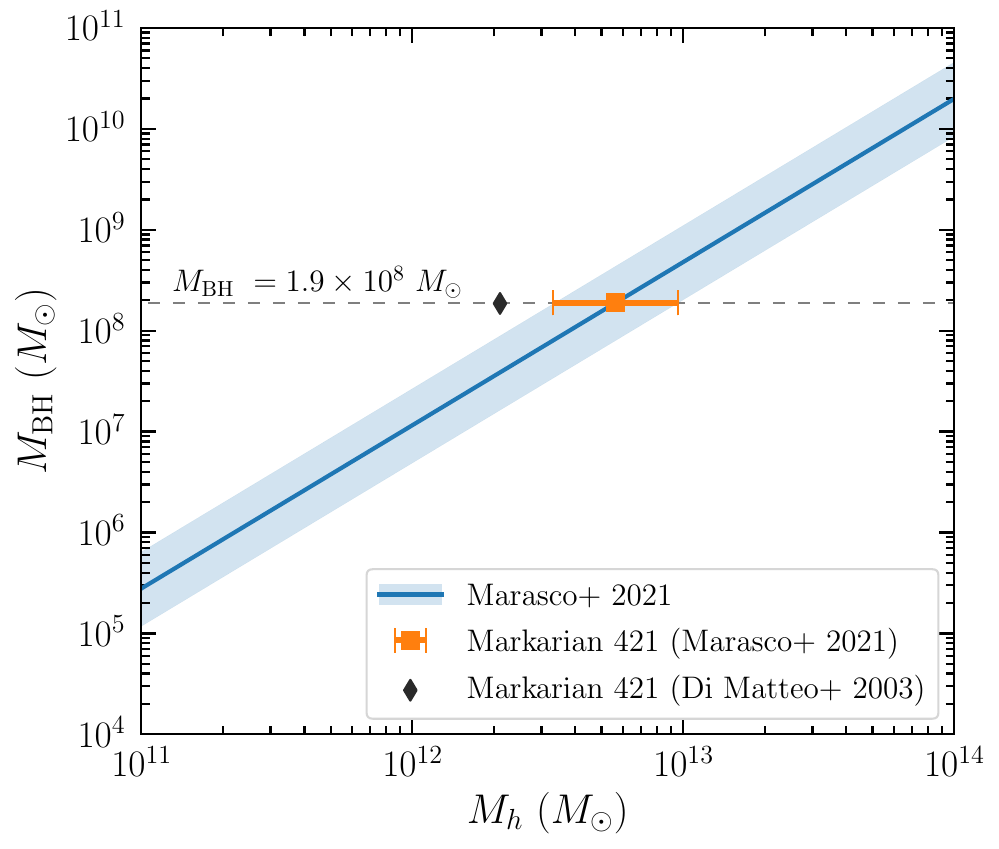} 
    \caption{\label{fig: MBH-Mhalo relation} The relation between the mass of the DM halo and the BH mass based on the observations of Ref.~\cite{Marasco:2021pkl}. We highlight the DM halo mass for the BH mass of \mkn with a square orange marker, and its value using the approach of Ref.~\cite{DiMatteo:2003zx} with a black diamond.}
\end{figure}

In Fig.~\ref{fig: MBH-Mhalo relation} we plot the $M_{\rm BH}-M_{\rm halo}$ relation based on the observations of Ref.~\cite{Marasco:2021pkl}. In case of \mkn, which has a BH mass of $M_{\rm BH} = 1.9 \times 10^8~M_\odot$~\cite{Barth:2002ac}, we get $M_{\rm halo} = 5.7 \times 10^{12}~M_\odot$ (orange square). 
Previous analyses have instead used the empirical equation of Ref.~\cite{DiMatteo:2003zx} which in our case implies $M_{\rm halo} = 2.1 \times 10^{12}~M_\odot$ (black diamond). Having calculated the mass of the halo, which is equal to $M_{200}$, we can find the value of the concentration based on the simulations of Ref.~\cite{Ishiyama:2020vao}. For the case of \mkn, the integrated value from the simulations is $c_{200} = 5.99$, that results in $r_s = 69.63~{\rm kpc}$ and $\rho_s = 1.65\times 10^{6}~M_\odot\,{\rm kpc^{-3}}$.

\begin{figure}[t!]
    \includegraphics[width=1\columnwidth]{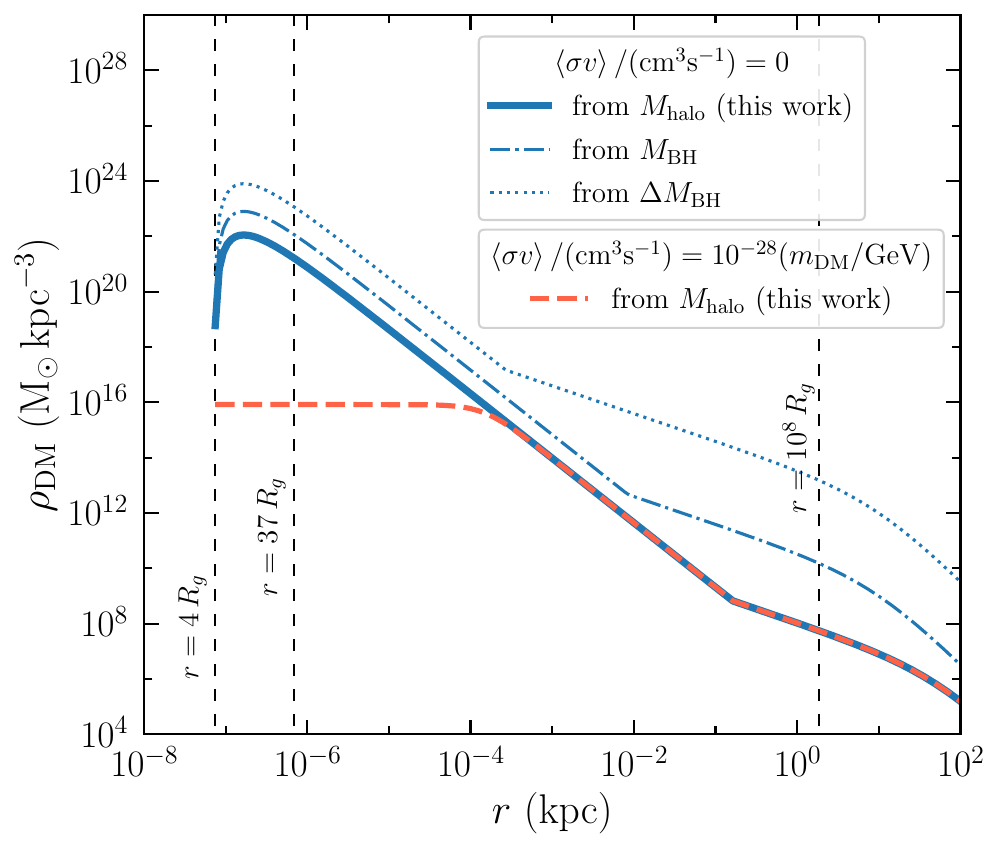} 
    \caption{\label{fig:DMprofile} The DM profile of \mkn based on Eq.~\eqref{equation: dark matter profile} and for the normalization values as given in the legend. We also include in the legend the approach followed to derive the normalization, namely whether the entire DM halo, the mass of the BH or the uncertainty of the mass of the BH is accounted for. With blue lines we show all the DM profiles in the case of a non-self-annihilating DM, and with a dashed red line, we show the DM profile with self-annihilation 
    $\left< \sigma v \right> /m_{\rm DM} = 10^{-28}\, \rm cm^3\, s^{-1}\, GeV^{-1}$ and using the DM halo for the normalization.}
\end{figure}

In Fig.~\ref{fig:DMprofile}, we show the DM profile defined in Eq.~\eqref{equation: dark matter profile} as a function of the radial distance from the BH as obtained in the present work considering non-self-annihilating DM particles (solid blue line) and self-annihilating DM particles with $\left<\sigma v \right> / ({\rm cm^3\,s^{-1}}) =  10^{-28}~(m_{\rm DM}/{\rm GeV})$ (dashed red line). We also plot for comparison the DM profile for non-self-annihilating DM particles which is obtained according to two different approaches adopted in previous works~\cite{Lacroix:2015lxa, Lacroix:2016qpq, Wang:2021jic, Granelli:2022ysi, Herrera:2023nww}. Namely, assuming 
a constant $r_s=10\,$kpc, we derive the normalization $\rho_s$ from integrating the DM mass up to the spike radius $R_{\rm sp}$ defined in Eq.~\eqref{eq:Rspike}, and then equating this mass to the BH mass $M_{\rm BH}$ (dot-dashed blue line) or its uncertainty $\Delta M_{\rm BH} = 1.2 \times 10^8~M_\odot$ (dotted blue line). We find $\rho_s= 3.95\times 10^{12}~M_\odot\,{\rm kpc^{-3}}$ and $\rho_s= 3.95\times 10^{9}~M_\odot\,{\rm kpc^{-3}}$, respectively.
The difference between the peak of the calculated DM profile following the proper approach and the approximated one is of the order of $\sim$ 1/50. Since the number density of DM particles scales linearly with the estimated CR cooling, the resulting upper limits in the CR-DM cross section may be overestimated by a such a factor. 

\end{document}